\documentclass{JHEP3}

\usepackage{latexsym,amsmath,amsfonts,amssymb}
\usepackage{multirow}
\usepackage[dvips]{graphicx}
\usepackage{mathrsfs,mathtools}

\newcommand{\tr}[1]{\text{Tr}\big[#1\big]}

\renewcommand{\d}[1]{\delta^d\Big(#1\Big)}

 \def\ii{{\rm i}}

\newcommand{\sect}[1]{\setcounter{equation}{0}\section{#1}}

\preprint{QMUL-PH-08-17\\arXiv:0810.0499}

\title{Operator Mixing and the AdS/CFT correspondence.}

\author{George Georgiou$^{a}$, Valeria L. Gili$^{a}$, Rodolfo Russo$^{a}$ \\
$^{a}$~Centre for Research in String Theory\\
Department of Physics, Queen Mary, University of London\\
Mile End Road, London E1 4NS, UK\\
E-mail: \email{G.Georgiou@qmul.ac.uk},
\email{V.Gili@qmul.ac.uk},
\email{R.Russo@qmul.ac.uk}}

\abstract{We provide a direct prescription for computing the mixing
  among gauge invariant operators in ${\cal N}=4$ SYM. Our approach is
  based on the action of the superalgebra on the states of the theory
  and thus it can be also applied to resolve the mixing in the dual
  string description. As an example, we focus on the supermultiplet
  containing the BMN operators with two impurities. On the field
  theory side, we derive the leading planar quantum corrections to the
  naive expression of the highest weight state. Then we use the same
  prescription in the BMN limit of the AdS$_5 \times S^5$ string
  theory and derive the form of the 2-impurity highest weight state.
  The string expression matches nicely the SYM result and provides a
  prediction for the mixing due to higher order quantum corrections in
  field theory.}


\begin{document}

\sect{Introduction}\label{Intro}

The operator mixing is an important aspect of any quantum field
theory. In the ${\cal N}=4$ Super Yang-Mills theory the mixing of
gauge invariant operators is strictly connected to the superconformal
properties of the theory. In fact, conformal symmetry puts strong
constraints on the form of two and three point correlators. However,
in concrete examples, these constraints are satisfied only if the
appropriate form of the operator is used. Actually, as it has been
extensively discussed in the
literature~\cite{Intriligator:1999ff,Bianchi:2002rw,Arutyunov:2002rs,Beisert:2002bb,Constable:2002vq,Eden:2003sj,Bianchi:2003eg},
this observation provides a concrete way to resolve the mixing and
obtain an explicit expression for the true eigenstates of the
dilatation operator. One starts by computing correlators among a set
of (classical) operators with the same naive scaling dimension and the
same quantum numbers. Requiring that these correlators take the form
dictated by conformal invariance implies a redefinition (mixing) of
the original naive basis for the set of operators considered. The most
common approach along these lines is to compute the 2-point functions
among the states considered and then look for an orthonormal basis.

${\cal N}=4$ SYM is characterized by two dimensionless parameters, the
rank of the gauge group $N$ and the 't~Hooft coupling $\lambda=g^2 N$.
Thus, we have two types of operator mixing: that between single and
multi trace operators governed by the $1/N$ expansion\footnote{In the
  BPS sector the 2-point correlators do not receive quantum
  corrections and the conformal invariance yields less stringent
  constraints. In the context of the AdS/CFT correspondence, it is
  anyway important to find an orthonormal basis for the BPS operator,
  which gives rise to an interesting combinatorial problem,
  see~\cite{Brown:2008ij,Kimura:2008ac} and reference therein.}, and
the mixing due to planar quantum corrections which can be
perturbatively computed as a series in $\sqrt{\lambda}$ (of course, in
general, the two types of corrections can combine and give terms
suppressed both by $\sqrt{\lambda}$ and $1/N$).  With the discovery of
integrable structures in $\mathcal{N}=4$ SYM~\cite{Minahan:2002ve}, it
was shown that, actually, the problem of diagonalizing the planar
dilatation operator is equivalent to finding the spectrum of the
Hamiltonian of a quantum spin chain. Nowadays, the complete
Hamiltonian is known up to one loop (order $g^2$) and Bethe Ansatz
techniques had been used to all orders in perturbation theory for
particular subsectors.

In this paper we present a different approach to the operator mixing
issue which directly relies on the ${\cal N}=4$ superalgebra. We start
from the well known statement that each operator in a supermultiplet
is annihilated by some of the (conformal) supercharges. For instance,
a non-BPS highest weight state $O$ is annihilated by all
superconformal charges ($S,~\bar{S}$) and none of the standard
supersymmetries ($Q,~\bar{Q}$). Schematically we have
\begin{equation}\label{si}
[S,O(x=0)]=0~~,~~\mbox{and}~~~~~ [Q,O(x=0)]\not=0~. 
\end{equation}
At the classical level, one can easily implement this requirement by
using the standard variations of the elementary fields composing the
gauge invariant operator. The easiest way to promote this approach to
the quantum level is to study the Ward identities of the
supersymmetric currents. For instance, the first equation
in~\eqref{si} can be rewritten as
\begin{equation}\label{sift}
\frac{\partial}{\partial y^\mu} \langle  S^{\mu}(y)\, O_1(x_1)\,
O(0)\rangle = -\ii\, 
\langle\delta_S O_1(x_1)\,O(0)\rangle\, \delta^4(y-x_1)~,
\end{equation}
where $S^\mu$ is the current related to the conformal supercharge $S$,
$O_1$ is an arbitrary operator and $[S,O_1]=\hat{O}_1$.  The fact that
$O$ is annihilated by $S$ translates into the absence in~\eqref{sift}
of a term proportional to $\delta(y)$. Of course~\eqref{sift} must
hold also beyond the tree-level approximation and we show that the
explicit computation of the quantum corrections can be used to resolve
the operator mixing.

In principle this approach can be applied to compute both the planar
and the $1/N$ corrections to the naive form of the non-BPS gauge
invariant operators. However, in our explicit example we focus only on
the planar mixing; in particular we derive the leading quantum
corrections to the highest weight state of the supermultiplet studied
in~\cite{Beisert:2002tn}. This state is a generalization of the usual
Konishi operator and has classical conformal dimension $J+2$
transforming in the $[0,J,0]$ representation of the $SU(4)$
R-symmetry. As it was suggested in~\cite{Bianchi:2003eg}, we find
that, at order $\sqrt{\lambda}$, the true highest weight state is a
combination of the naive form containing only scalar fields and a
correction term with two fermion impurities.  This is reminiscent of
the mixing discussed in~\cite{Green:2005ib} in the context of
instanton corrections, but is new from the quantum spin chain
approach. As we will discuss at the end of section \ref{FT}, this
mixing between scalar and fermion impurities is not captured by the
one-loop Hamiltonian, even if it appears at order $\sqrt{\lambda}$.
The situation is similar to the single/double trace mixing discussed
in~\cite{Beisert:2002bb,Constable:2002vq}: the form of the operators
at a certain order in perturbation theory requires the knowledge of
the dilatation operator at higher orders.

A nice feature of the method discussed here is that it can also be
applied directly on the string side of the AdS/CFT correspondence,
thus shedding some light on how the operator mixing is realized in the
dual string description. There is a one-to-one map between the field
and the string theory superalgebra~\cite{Maldacena:1997re} and so we
can look for the string states that are annihilated by the same
supercharges used in the field theory computation. This approach
provides a natural dictionary between the spectra of the two
descriptions.  Of course, its implementation in the full type IIB
superstring theory on AdS$_5 \times S^5$ is beyond reach right now.
However, we can easily carry out this computation in the BMN
limit~\cite{Berenstein:2002jq}, where we focus only on states with a
large $U(1)$ R-charge $J$. In this limit, type IIB string theory in
the light cone is described by a free 2-dimensional world-sheet
Lagrangian and the expression of all supercharges is explicitly
known~\cite{Metsaev:2001bj,Metsaev:2002re}. We again consider the
sector of 2-impurity states and derive the highest weight state of the
multiplet in the free PP-wave string theory. The string expression
matches the large $J$ limit of the perturbative SYM result and
provides a prediction for the higher order quantum corrections in
field theory. However, the numerical agreement between the string and
the field theory results is somehow surprising. BMN string theory is a
reliable approximation of the full $AdS_5\times S^5$ theory in the
limit $\lambda,J\rightarrow\infty$, with $\frac{\lambda}{J^2} =
\lambda'$ fixed, which is far away from the standard perturbative
field theory ($\lambda \ll 1$). We extend the string result at weak
coupling by supposing the validity of the BMN scaling. This is known
to break down at four loops in gauge theory~\cite{Eden:2006rx}. Thus,
it is unlikely that this numerical agreement between the large $J$
limit of the form of the operator and the small $\lambda$ behavior of
the string expression survives at higher loops.

The structure of the paper is the following. In Section~\ref{ST} we
focus on free IIB string theory in the BMN limit. We summarize how the
superalgebra is realized and we use it to build the highest weight
states with two impurities. In the strong curvature limit, one
recovers the usual expression which contains only scalar impurities
and is directly connected to free SYM expression for the primary
operators. The exact expression for the string highest weight states
involves also fermion and scalar impurities suggesting a precise
mixing pattern on the SYM side of the correspondence. In
Section~\ref{FT} we focus on the SYM side of the correspondence. By
studying the supersymmetric Ward identities we give a perturbative
derivation of the action of the supercharges up to order $g$ on gauge
invariant operators with scalar fields. In~\cite{Beisert:2003ys} this
result was derived by using exclusively the $SU(2|3)$ subgroup of the
${\cal N}=4$ symmetry algebra, together with just one dynamical input
such as the known anomalous dimension of the Konishi operator. The
action of the supersymmetry generators has been
studied~\cite{Kim:2002if,Aniceto:2008wi} also in the plane-wave matrix
model in a spirit similar to what is done in this paper. In
Section~\ref{sec:comp} we compare the strong curvature expansion of
the two impurity string states derived in Section~\ref{ST} against the
1-loop corrected field theory highest weight states in the large $J$
limit. The two results agree by using the standard BMN/SYM dictionary
and this suggests that at order $g^2$ the field theory primary
operators should contain also space-time derivatives. In the
Conclusions we discuss some possible developments where the results
presented in this paper can play an important role. Two Appendices
contain our conventions for the ${\cal N}=4$ SYM theory and all
technical results useful for the derivation discussed in
Section~\ref{FT}.

\sect{Operator mixing in (BMN) string theory}\label{ST}

Let us consider type IIB string theory on the maximal supersymmetric
PP-wave background~\cite{Blau:2001ne}. In the light-cone gauge this
theory is described by a free two dimensional action. So upon
quantization we have eight towers of bosonic and fermionic harmonic
oscillators ($a^\dagger_n$, $b^\dagger_n$)
transforming in the vector and spinor representation respectively of
the $SO(4)\times SO(4)$ subgroup of the full $SO(2,4)\times SO(6)$
isometry group of AdS$_5 \times
S^5$~\cite{Metsaev:2001bj,Metsaev:2002re}. The physical spectrum is a
subset of the Fock space generated by these creations operators which
consists of all states satisfying the level matching
condition\footnote{We follow the notations of~\cite{DiVecchia:2003yp},
with some small changes so to have a string theory supersymmetry
algebra that agrees with the field theory one following from the
conventions of Appendix~\ref{sec:GTconv}.}
\begin{equation} \label{Tdef}
T|s\rangle=0~~,~~~~\mbox{with}~~~
T = \sum_{n=1}^\infty {n}
\left[(b_n^\dagger b_{-n} + b_{-n}^\dagger b_n) -
\ii (a_n^\dagger a_{-n} - a_{-n}^\dagger a_n) \right]~,
\end{equation}
where we suppressed all space-time indices that are contracted in the
standard $SO(4)\times SO(4)$ invariant way. The light-cone Hamiltonian
is
\begin{equation}
  H = \frac{1}{\mu\alpha' |p^+|} \sum_{n= -\infty}^{\infty} {\omega_n}
  \left[ a_{n}^{\dagger} a_n + b_{n}^{\dagger} b_n \right]~,
\label{hami89}
\end{equation}
where $\omega_{n} = \sqrt{n^2 + (\mu \alpha' p^+)^2}$, $p^+$ is the
light cone momentum of the state and $\mu$ is the parameter setting
the curvature of the PP-wave background (in the following we will
define, as usual, $\alpha \equiv \alpha' p^+$).  This background
preserves 32 supercharges. Half of them are purely kinematical and
contain only the zero-mode oscillators
\begin{equation}
  \label{eq:qplus}
  Q^+ = \sqrt{2 |\alpha|} \left[e(\alpha)\mathbb{P}^- b_0 +
    \mathbb{P}^+ b_0^\dagger\right]~,~~ 
  \bar{Q}^+ =  \sqrt{2|\alpha|} \left[\mathbb{P}^-
    b^\dagger_0 + e(\alpha) \mathbb{P}^+ b_0 \right]~,
\end{equation}
where $\mathbb{P}^{\pm} = (1\pm \Pi)/2$ and $\Pi$ is the appropriate
$16 \times 16$ block of the matrix $\prod_{i'=1}^4 \Gamma^{i'}$, where
the index $i'$ is restricted to the flavor $SO(4) \subset SO(6)$ and
the $\Gamma$'s ($\gamma$'s ) indicate the $SO(1,9)$ ($SO(8)$) Gamma
matrices respectively. The remaining sixteen supercharges are
dynamical and display a non-trivial dependence on $\mu\alpha$
\begin{eqnarray}\label{qminuso}
Q^- &=& e(\alpha)\sqrt{\frac{1}{2}} \gamma \left[a_0
  \left(1+e(\alpha) \Pi\right)b_0^\dagger + e(\alpha) a_0^{\dagger}
  \left(1-e(\alpha) \Pi\right) b_0\right] + \\ &+&
  \frac{1}{\sqrt{|\alpha|}} \sum_{n=1}^{\infty} \sqrt{n} \gamma
  \left[a_n^{\dagger} P_n b_{-n} + e(\alpha) a_n P_n^{-1}
  b_n^{\dagger} + \ii a_{-n}^{\dagger} P_n b_n - \ii e(\alpha) a_{-n}
  P_n^{-1} b_{-n}^{\dagger} \right]\;,\nonumber \\ \nonumber
\bar{Q}^-&=& \sqrt{\frac{1}{2}} \gamma \left[e(\alpha) a_0
  \left(1-e(\alpha) \Pi\right)b_0^\dagger +a_0^{\dagger}
  \left(1+e(\alpha)\Pi\right) b_0\right] + \\
&+& \frac{1}{\sqrt{|\alpha|}} \sum_{n=1}^{\infty}
\sqrt{n} \gamma \left[a_n^{\dagger} P_n^{-1} b_n
  + e(\alpha) a_n P_n b_{-n}^{\dagger} + \ii a_{-n}^{\dagger} P_n^{-1}
  b_{-n} - \ii e(\alpha) a_{-n} P_n b_{n}^{\dagger} \right]\;,\nonumber
\end{eqnarray}
where $e(\alpha)=1$ if $\alpha>0$ while $e(\alpha)=-1$ if $\alpha<0$
and
\begin{equation}
  \rho_{n} =\frac{\omega_n -n}{\mu \alpha}
  ~~,~~ 
  P_{n}^{\pm 1} =
  \frac{1}{\sqrt{1-\rho^2_{n}}} (1  \mp \rho_{n} \Pi)~.
\label{cppm}
\end{equation}

The standard prescription~\cite{Berenstein:2002jq} for building the
dictionary between string and field theory states is to identify the
creation modes with the presence of ``impurities'' (i.e. fields with
$\Delta-J=1$) in the corresponding gauge theory operator. Thus the
dictionary is usually set-up at the level of the basic constituents
(letters) by checking that they transform in the same way under the
$SO(4)\times SO(4)$ symmetry of the problem. For instance, the
relation between the gauge and string theory expression for the $SO(4)
\times SO(4)$ singlet with two scalar impurities is usually written as
follows\footnote{See Appendix~\ref{sec:GTconv} for our field theory
conventions.}
\begin{equation}\label{usst}
   \sum_{p=0}^J \cos{\frac{\pi n (2 p + 3)}{J + 3}}
  \tr{ \Phi_{AB} Z^p \Phi^{AB} Z^{J-p}}
  \,\longleftrightarrow\,\sum_{i'=1}^4
    {\alpha^\dagger}_n^{i'}
    {\alpha^\dagger}_{-n}^{i'}
  |\alpha\rangle~,
\end{equation}
where $|\alpha\rangle$ is the vacuum state of fixed light-cone
momentum $p^+$ and $\alpha^\dagger_n$ ($\alpha^\dagger_{-n}$) are the
oscillators creating left (right) moving excitations on the
string\footnote{In all other formulae of this paper we use the string
field theory oscillators $a_n$, which are related to the $\alpha_{\pm
n}$ as follows: $\alpha_{\pm n} = \frac{1}{\sqrt{2}}(a_n \mp \ii
a_{-n})$.}.

Here we will follow a different approach to the construction of the
field/string theory dictionary. We first identify the supersymmetry
generators in the two descriptions by requiring that they satisfy the
same algebra. Then, we derive the highest weight states of the string
and field theory algebra separately. The first entry of the dictionary
between the two spectra just consists in relating the two highest
weight states. Then it is straightforward to build the dictionary for
the whole supermultiplet: we just need to act on the the highest
weight state in each description with supercharges that have been
already identified. The two approaches yield the same dictionary
between the string and the field theory spectra in the large
$\mu\alpha$ limit. What is more surprising is that even the first
subleading corrections agree, as we will see in
section~\ref{sec:comp}.

By comparing the string and field theory superalgebras, one obtains
(for $\alpha>0$) the following correspondence
\begin{equation}
  \label{eq:algidQ}
  Q_{\alpha,A=1,2} \leftrightarrow \mathbb{P}^+ Q^+ ~,~~
  Q_{\alpha,A=3,4} \leftrightarrow \mathbb{P}^+ Q^- ~,~~
  \bar Q^{\dot\alpha,A=1,2} \leftrightarrow \mathbb{P}^- \bar{Q}^- ~,~~
  \bar Q^{\dot\alpha,A=3,4} \leftrightarrow \mathbb{P}^- \bar Q^+ ~,
\end{equation}
where $Q_\alpha$ and $\bar Q^{\dot\alpha}$ are the standard gauge
theory supercharges~\eqref{eq:4dcurrmu}, and $Q^\pm$, $\bar Q^\pm$ are
the supersymmetry operators in the PP-wave string
theory~\eqref{eq:qplus}--\eqref{qminuso}. In the BPS sector a highest
weight state is annihilated also by half of the transformations in
\eqref{eq:algidQ}. In fact, on the field theory side the operator
$\tr{Z^J}$ is invariant under transformation generated by $Q_{3,4}$
and $\bar Q^{1,2}$ and the same is true for the corresponding string
state $|\alpha\rangle$.  The other sixteen string supercharges
correspond to the superconformal symmetries of the gauge theory
description, see~\eqref{supc}
\begin{equation}
  \label{eq:algidS}
  S^{A=1,2}_{\alpha} \leftrightarrow \mathbb{P}^+ \bar Q^+ ~,~~
  S^{A=3,4}_{\alpha} \leftrightarrow \mathbb{P}^+ \bar Q^- ~,~~
  \bar S_{A=1,2}^{\dot\alpha} \leftrightarrow \mathbb{P}^- Q^- ~,~~
  \bar S_{A=3,4}^{\dot\alpha} \leftrightarrow \mathbb{P}^- Q^+ ~.
\end{equation}
Thus any highest weight states should be annihilated by all operators
in~\eqref{eq:algidS}
\begin{equation}
  \label{eq:stsi}
  \mathbb{P}^+ \bar Q^+ |hws\rangle= \mathbb{P}^+ \bar Q^- |hws\rangle=
  \mathbb{P}^- Q^-  |hws\rangle= \mathbb{P}^- Q^+ |hws\rangle= 0~. 
\end{equation}
Then it is clear that we should focus on the string states that do not
contain any $b_0^\dagger$ so that they are annihilated by $\mathbb{P}^+
\bar Q^+$ and $\mathbb{P}^- Q^+$. The conditions following from the
remaining supercharges must be solved case by case: here we will
consider the multiplets containing the states with two string creation
operators and show that the 2-impurity highest weight states are not
given simply by the Eq.~\eqref{usst}.

The first observation is that the string state in~\eqref{usst} is
annihilated by the dynamical supercharges in~\eqref{eq:algidS} {\em
only} in the $\mu\alpha \to\infty$ limit. In fact, when we compute the
$\mathbb{P}^\pm$ projections of the dynamical supercharges
in~\eqref{eq:algidS}, we have to separate the terms with a Gamma
matrix in the ``flavor'' $SO(4)$ (indicated with an index $i'$) from
those with a Gamma in the ``space-time'' $SO(4)$ (indicated with
$i$). $\Pi$ commutes with $\gamma^{i'}$ and anticommutes with
$\gamma^i$. Thus, for instance, in the case $\alpha>0$ we have that
the string charges corresponding to $S^{A=3,4}_{\alpha}$
are
\begin{eqnarray}\label{q-proj}
\mathbb{P}^+ \bar{Q}^- &=& \sqrt{2}
  \left[\gamma^{i} a_0^{i} \mathbb{P}^- b_0^\dagger + \gamma^{i'}
    a_0^{i'\,\dagger} \mathbb{P}^+ b_0\right] +
  \frac{1}{\sqrt{\mu|\alpha|}} \sum_{n=1}^{\infty} \sqrt{n}
  \Bigg\{ \\ & &
  \gamma^{i'} \left[a_n^{i'\dagger} \mathbb{P}^+ b_{n} + \ii
    a_{-n}^{i'\dagger} \mathbb{P}^+ b_{-n}\right] U_n^{-\frac 12} +
   \gamma^{i'} \left[a_n^{i'} \mathbb{P}^+ b_{-n}^\dagger - \ii
    a_{-n}^{i'} \mathbb{P}^+ b_{n}^\dagger \right] U_n^{\frac 12}
  \nonumber  \\ &+& \nonumber
  \gamma^i \left[a_n^i \mathbb{P}^- b_{-n}^{\dagger} -\ii
   a_{-n}^i \mathbb{P}^- b_n^{\dagger}\right] U_n^{-\frac 12}
  + \gamma^i \left[a_n^{i \dagger} \mathbb{P}^- b_{n}
  + \ii a_{-n}^{i \dagger} \mathbb{P}^- b_{-n} \right] U_n^{\frac 12}
\Bigg\}~,
\end{eqnarray}
where $U_n^{\pm 1} \equiv \frac{1\mp\rho_{n(1)}}{1\pm\rho_{n(1)}}$ and
the repeated indices are summed. It is interesting to consider the
form of the dynamical supercharges in the large $\mu\alpha$ limit. In
this limit we have that $U_n \sim n/(2\mu \alpha)$, so the terms with
$U^{-1/2}_n$ dominate over those with $U_n^{1/2}$ and the leading
contribution to~\eqref{q-proj} is schematically $\mathbb{P}^+ \bar{Q}^-
\sim \gamma^{i} a^{i} \mathbb{P}^- b^\dagger$ along the space-time
directions and $\mathbb{P}^+ \bar{Q}^- \sim \gamma^{i'} a^{i'\,\dagger}
\mathbb{P}^+ b$ along the flavor directions. This result can be matched
directly against the $g\to 0$ form of the field theory superconformal
transformation of Appendix~\ref{sec:GTconv}: $S \psi(0) \sim Z(0)$ and
$S \partial Z(0) \sim \sigma \psi(0)$, where again we have suppressed
all numerical factors and indices. In the same way, the large
$\mu\alpha$ limit of $\mathbb{P}^+ Q^-$ and $\mathbb{P}^- \bar Q^-$ agrees
with the gauge theory supersymmetry transformations at $g=0$, as
summarized by the dictionary~\eqref{eq:algidQ}.

Let us now consider the action of~\eqref{q-proj} on the string state
in~\eqref{usst}. It is clear that the second term of the second line
annihilates this state only in the large $\mu\alpha$ limit, showing
that it is not an exact highest weight state of the string
superalgebra. This suggests that also on the field theory side the
operator in~\eqref{usst} is a superconformal primary only in the
$g_{YM} \to 0$ limit. On the string side it is clear how to modify the
state in~\eqref{usst} so to find the true highest state weight of the
multiplet. We need to add a contribution which contains two fermionic
oscillators and is not annihilated by the first term of the second
line in~\eqref{q-proj}. The coefficient is chosen to
satisfy~\eqref{eq:stsi}. By repeating the same procedure also for
$\mathbb{P}^- Q^-$, one obtains that the 2-impurity states
satisfying~\eqref{eq:stsi} exactly are
\begin{equation}
\label{shws}
|n\rangle = \frac{1}{4(1+U_n^2)}
\left[
    {a^\dagger}_n^{i'}
    {a^\dagger}_n^{i'}
    \,+\,
    {a^\dagger}_{-n}^{i'}
    {a^\dagger}_{-n}^{i'}
  + 2 U_n 
  b_{-n}^\dagger\, \Pi\; b_n^\dagger
  - U_n^2 \left(
    {a^\dagger}_n^{i}
    {a^\dagger}_n^{i}
    \,+\,
    {a^\dagger}_{-n}^{i}
    {a^\dagger}_{-n}^{i}
  \right)\right]
  |\alpha\rangle,  
\end{equation}
where the overall normalization has been fixed in such a way that the
state is normalized to one: $\langle n|n \rangle=1$. The main feature
of~\eqref{shws} is the mixing between various types of impurities. At
leading order in the $\mu\alpha\to\infty$ expansion, we have only
scalar impurities (${a^\dagger}_n^{i'}$). The first corrections appear
at order ${\cal O}(1/(\mu\alpha))$ and are quadratic in the fermionic
oscillators. According to the standard dictionary between the PP-wave
and the field theory parameters, this translates into a quantum
correction of order $\lambda'$.  At the next order (${\cal
  O}(1/(\mu\alpha)^2)$) also vector impurities appear and we expect
that the same pattern is present also in field theory. Finally, let us
stress again that, starting from the state in~\eqref{shws}, it is
tedious but straightforward to build the whole supermultiplet by using
the supercharges in~\eqref{eq:algidQ}.

\sect{Operator mixing in field theory}\label{FT}

The analysis of the previous section has shown that in general the
string theory highest weight states involve mixing between different
kinds of impurities. It would be desirable to see the same pattern
appearing in perturbative field theory. In this section, we evaluate
the first correction to the classical form of the highest weight
operator involving two impurities and find perfect agreement, in the
appropriate limit, with the string theory expression~\eqref{shws}.  We
follow the same approach discussed in the string theory context and
compute the form of the highest weight state by looking for field
theory operators that satisfy~\eqref{si}.

As already mentioned in the previous section, the gauge theory
supersymmetry transformations at $g=0$ agree with $\mu\alpha\to\infty$
limit of the superstring ones and the scalars are annihilated by all
$S$ and $\bar{S}$~\eqref{supc}. Thus any composite operator built solely
from scalars is a primary state at leading order, and the 2-impurity
SYM primaries are the operators already introduced in~\eqref{usst}:
\begin{equation}\label{ng}
  \mathcal{O}^{(0) J}_n = \sqrt{\frac{N_0^{-J-2}}{(J + 3)}}
  \sum_{i=1}^3
    \sum_{p=0}^J 
    \cos{\frac{\pi n (2 p + 3)}{J + 3}}\tr{Z_i Z^p \bar Z_i
  Z^{J-p}}~, 
\end{equation}
where the normalization $N_0=N/(8\pi^2)$ is fixed to have\footnote{The
  factor of $(-1)^{J+2}$ disappears after rotating to Euclidean
  spacetime.} $ \langle \bar{\mathcal{O}}^{(0) J}_n\!(x)\;
\mathcal{O}^{(0) J}_n\!(0)\rangle = \frac{(-1)^{J+2}}{(x^2)^{J+2}}$.

Things change if one considers the full interacting quantum theory.
In this case, most of the aforementioned states are not annihilated by
all the superconformal charges and the true primaries are not built
with scalar impurities only. For instance, at first order in $g$, we
have
\begin{equation}
\label{s1}
\bar{S}_A^{\dot\alpha} \Phi_{BC}
\Phi_{DE}(0)=  \ii\, \frac{g N}{32 \pi^2} \left(
\epsilon_{ABC[D} \bar\psi_{E]}(0) -\epsilon_{ADE[B}
  \bar\psi^{\dot\alpha}_{C]}(0) \right)~,
\end{equation} 
where $\epsilon_{ABC[D} \bar\psi_{E]}=\frac{1}{2}
(\epsilon_{ABCD} \bar\psi_{E}-\epsilon_{ABCE} \bar\psi_{D})$. 
If we restrict the indices to the $SU(2|3)$ sector, this expression
agrees\footnote{One has to take into account that our supersymmetry
  algebra agrees with that of~\cite{Dolan:2002zh}, where the
  (super)conformal generators $S$ and $K$ are normalized in a
  different way from~\cite{Beisert:2003ys}.} with that of
\cite{Beisert:2003ys}. Here we give a diagrammatic derivation of this
result that immediately leads to the $SU(2,2|4)$ form of~\eqref{s1}.
The relevant field theory diagrams are depicted in 
figures~\ref{fig:oneloop} and \ref{fig:comm}, 
where the classical form of the
superconformal transformation is combined with a Yukawa coupling.

\FIGURE[!t]{
  \centering
  \includegraphics[width=0.4\textwidth]{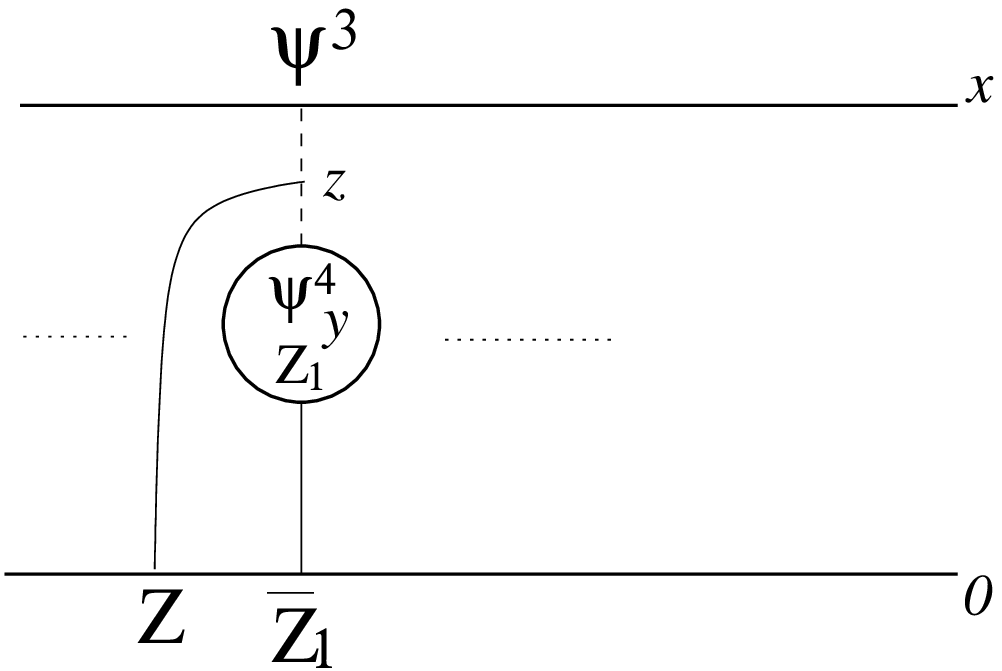}
\,\,
  \includegraphics[width=0.4\textwidth]{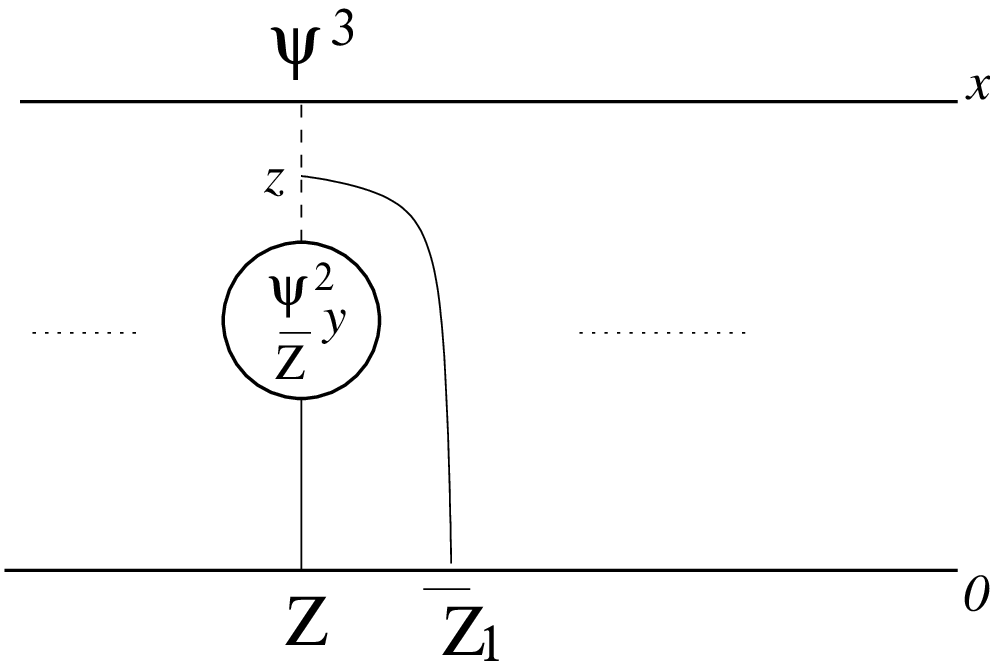}
  \caption{Diagrams contributing in the one loop calculation of \eqref{3point}.
The solid lines denote scalar propagators while the dashed 
ones fermion propagators.}
  \label{fig:oneloop}}

For sake of concreteness, let us focus on the action of $\bar S_1$ on
the scalar fields $Z \bar Z_1$; we compute the 3-point function
\begin{equation}\label{3point}
  (G_3^{\mu})^{i\,k}_{j\,l}=\langle \bar{S}^{\mu\dot\alpha}_1(y)\,
  (\psi^3_{\gamma})^i_j(x)\,
  (Z\bar{Z}_1)^k_l (0)\rangle~
\end{equation}
and demand that it is compatible with~\eqref{s1}. In equation
\eqref{3point} we have written explicitly the free indices of the
operators which are not scalars under the $SU(N)$ color group. The
3-point function of \eqref{3point} at order $g$ receives contribution
from the three terms in \eqref{eq:bscc}.  In particular, the
contributions related to the last and the penultimate term of
\eqref{eq:bscc} are depicted in fig.~\ref{fig:oneloop}
and~\ref{fig:comm} respectively.  Finally, the contribution of the
second term in fig.~\eqref{eq:bscc} can be obtained from the diagrams
of \ref{fig:oneloop} by remembering that there is an additional
derivative acting on the field $Z_1(y)$.  By using the propagators and
the vertices summarized in the Appendix~\ref{sec:GTconv}, it is
straightforward to see that the two diagrams of fig.~\ref{fig:oneloop}
yield the same integral. So we have
\begin{equation}\label{diag.(a)}
G^{(1)\mu}_3= -\,4\sqrt{2} 
\,\Delta(y)\,\frac{N}{2^4}\,(-\ii2\sqrt{2})\, g\!\! \int d^4z \, 2
\ii^2 \,\sigma^{\nu}_{\gamma\dot\beta}\partial^z_\nu\Delta(x-z)\,
\epsilon^{\dot\beta\dot\gamma}\,
\sigma^{\kappa}_{\alpha\dot\gamma}\partial^z_\kappa\Delta(y-z)\,
\bar{\sigma}^{\mu\dot\alpha \alpha}\, \Delta(z)\,.
\end{equation}
\FIGURE[!t]{
  \includegraphics[width=0.6\textwidth]{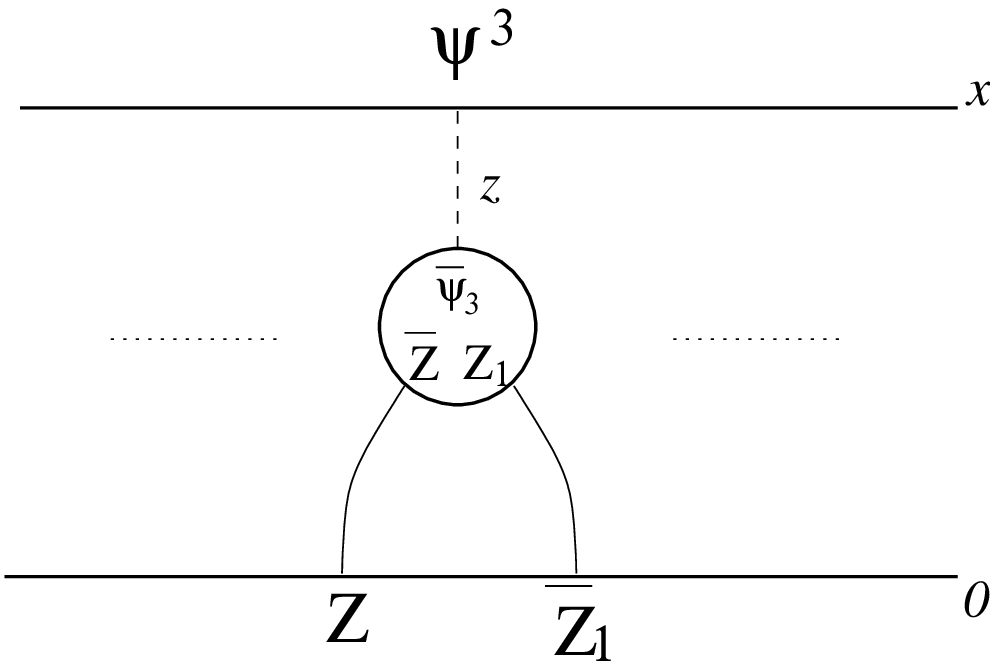}
  \caption{This is the diagram contributing to \eqref{diag2}}
  \label{fig:comm}} 

Some comments are in order: the factor of $4\sqrt{2}$ comes from the
current, of $-\ii 2\sqrt{2} g$ is due to the insertion of the Yukawa
coupling and the last factor of $2$ in the integral takes into account
the two diagrams in fig.~\ref{fig:oneloop}, while the overall sign
comes from the fermionic Wick contractions.  Finally, adopting the
conventions in~\eqref{colcon}, the color algebra gives, in the large
$N$ limit, the factor $\frac{N}{2^4} \delta^i_l\delta^k_j$. In
formula~\eqref{diag.(a)} and in what follows we drop the tensorial
$SU(N)$ structure, which is unnecessary for our computation, by
defining the quantity $G^\mu_3$ via the relation
$(G^\mu_3)^{i\,k}_{j\,l} = G^\mu_3 \delta^i_l\delta^k_j$.

By using the integral \eqref{integral}
one obtains from~\eqref{diag.(a)}:
\begin{equation}\label{(a)y=0}
G^{(1)\mu}_3 = - {2 g N}\,\Delta(y)
\,\bar{\sigma}^{\mu\dot\alpha \alpha}\, \sigma^{\nu}_{\gamma\dot\beta}
\epsilon^{\dot\beta\dot\gamma}\,\sigma^{\kappa}_{\alpha\dot\gamma}\,\frac{1}{(4
\pi^2)^2} \, \frac{y_\kappa x_\nu}{x^2 y^2(x-y)^2}~.
\end{equation}
The next step is to evaluate the diagram of fig.~\ref{fig:comm}. This
gives
\begin{eqnarray}\label{diag2}
G^{(2)\mu}_3 &=& 4 g \frac{N}{2^3} y_{\tau}
\bar{\sigma}^{\tau\dot\alpha \alpha}\, 
\sigma^{\mu}_{\alpha\dot\beta}\, \epsilon^{\dot\beta\dot\gamma}\,
[\Delta(y)]^2\, \sigma^{\nu}_{\gamma\dot\gamma}\,
\partial^x_\nu\Delta(x-y) \\ \nonumber
&=& \frac{g N}{16 \pi^2}\bar{\sigma}^{\tau\dot\alpha \alpha}\,
\partial^y_\tau\Delta(y)\,\sigma^{\mu}_{\alpha\dot\beta}\, 
\epsilon^{\dot\beta\dot\gamma}\,
\sigma^{\nu}_{\gamma\dot\gamma}\,
\partial^x_\nu\Delta(x-y)~.
\end{eqnarray}
The final ingredient we need is the contribution of the diagrams
coming from the second term of~\eqref{eq:bscc}. 
These give an expression very similar to that of the diagrams
of fig.~\ref{fig:oneloop}:
\begin{multline}
\label{diag3}
G^{(3)\mu}_3= - 2\sqrt{2}\,\frac{g N}{2^4}\,(-2\sqrt{2}\,\ii)\,
y_{\tau} \bar{\sigma}^{\tau\dot\alpha \alpha}\,\times 
\\\int d^4z 2
\ii^2 \,\sigma^{\nu}_{\gamma\dot\delta}\partial^z_\nu\Delta(x-z)\,
\epsilon^{\dot\delta\dot\gamma}\,
\partial^z_\kappa\Delta(y-z)\,
\sigma^{\kappa}_{\beta\dot\gamma}\Delta(z)
\sigma^{\rho}_{\alpha\dot\beta}
\bar{\sigma}^{\mu\dot\beta \beta}\, 
\partial^y_\rho\Delta(y) ~.
\end{multline}
After some algebra and by using~\eqref{integral} one gets:
\begin{equation}\label{diag3f}
G^{(3)\mu}_3= 2\, g N \Delta(y)\,
\bar{\sigma}^{\mu\dot\alpha \beta}\,
\frac{1}{(4\pi^2)^2}\frac{y_{\kappa} x_\nu}{x^2y^2(x-y)^2}\,
\sigma^{\nu}_{\gamma\dot\delta}\,\epsilon^{\dot\delta\dot\gamma}\,
\sigma^{\kappa}_{\beta\dot\gamma}~.
\end{equation}
By comparing~\eqref{diag3f} to~\eqref{(a)y=0} one can see that they
precisely cancel. We are now in position to write the final
expression for $G_3^{\mu}$.  This reads:
\begin{equation}\label{G3}
G_3^{\mu}=G_3^{(2)\mu}=\frac{
  gN}{16\pi^2}\bar{\sigma}^{\tau\dot\alpha \alpha}\, 
  \partial_\tau^y \Delta(y)\,\sigma^{\mu}_{\alpha\dot\beta}\,
    \epsilon^{\dot\beta\dot\gamma}\,\sigma^{\nu}_{\gamma\dot\gamma}\,
                \partial^x_\nu\Delta(x-y)~.
\end{equation}
It is now straightforward to find the divergence of \eqref{G3}.
\begin{equation}\label{div}
\partial_{\mu}^y G_3^{\mu}=-\frac{\ii g N}{16 \pi^2}(\delta^{(4)}(y)
 \epsilon^{\dot\alpha\dot\gamma}\,\sigma^{\nu}_{\gamma\dot\gamma}\,
\partial^x_\nu\Delta(x)+...)~,
\end{equation}
where the dots represent a term proportional to $\delta^{(4)}(x-y)$
of which we will make no use in what follows.
It is now straightforward to obtain the superconformal variation of 
the operator $ZZ_1(0)$ with respect to $\bar{S}_1^{\dot\alpha}$.
By comparing \eqref{div} to 
\begin{equation}\label{zeta1}
\partial_{\mu}^y\langle S^{\mu}(y)\,O_1(x)\,O(0)\rangle=-\ii \delta^4(x-y)
\langle\delta_SO_1(x)\,O(0)\rangle+
\ii\,\delta^4(y)\langle O_1(x)\,\delta_S O(0)\rangle 
\end{equation}
 one gets:
\begin{equation}\label{pre32}
\bar{S}_1^{\dot\alpha}Z \bar Z_1
=\frac{\ii g N}{8 \pi^2}\bar\psi_{3}^{\dot\alpha}.
\end{equation}
Notice that the different sign in front of the second term of the
r.h.s of eq. \eqref{zeta1} is due to the fermionic nature of the
operator $O_1$.  When the scalars are in the opposite ordering $\bar
Z_1 Z$, then the action of $\bar S_1$ is the same but for the overall
sign: $\bar{S}_1^{\dot\alpha}Z \bar Z_1=-\bar{S}_1^{\dot\alpha}\bar
Z_1 Z$. In fact from the form of the currents~\eqref{supc} and of the
Yukawa couplings~\eqref{n4l} it is clear that the diagrams
contributing to this two cases always have a relative minus sign.
Finally, when $\bar S_1$ acts on scalar of the same flavor, as in
$\bar{S}_1^{\dot\alpha}Z \bar Z$, there is an additional factor of
$1/2$, which again follows from the form of the
currents~\eqref{supc}. So finally, by using~\eqref{pre32} and these
observations, one arrives at the result~\eqref{s1}.

\FIGURE[!t]{
\includegraphics[width=0.6\textwidth]{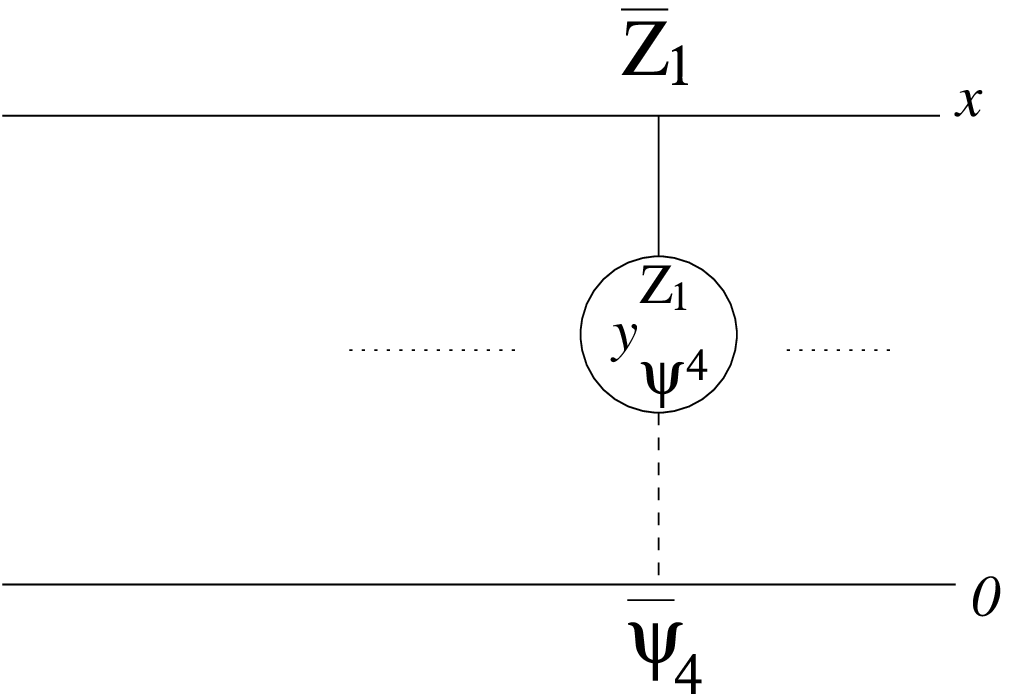}
  \caption{This diagram represent the classical variation \eqref{s2}.}
  \label{fig:treelevel}}

Of course, we can follow the same procedure in order to derive the
classical variations. For instance, at leading order in $g$ the action
of a conformal supersymmetry on an elementary fermion is given by the
diagram in fig.~\ref{fig:treelevel} which yields
\begin{equation}
\label{s2}
\bar{S}_A^{\dot\alpha} \bar{\psi}_{B \dot{\beta}}=
  4\sqrt{2}\ii \Phi_{AB}\delta^{\dot\alpha}_{\dot\beta}~, 
\end{equation}
where again all fields are at $x=0$

An independent check on the coefficient of equation \eqref{s1} can be
performed via the $SU(2,2|4)$ algebra by using only the well known
expression for the spin chain Hamiltonian at order
$g^2$~\cite{Minahan:2002ve}. If we act with the algebra on scalars of
different flavour, at order $g^2$ we have $
\left\{\bar{Q}^{A\dot{\alpha}}, \bar{S}^{\dot{\beta}}_B \right\} \,=\,
\epsilon^{\dot{\alpha}\dot{\beta}}\, \delta^A_B\, 2 \mathbb{H}$, with
$\mathbb{H} = \frac{g^2 N}{8\pi^2} \left( \mathbb{I}-\mathbb{P}
\right)$.  In particular, we can restrict ourselves to
\begin{equation}
  \label{eq:qs}
\left\{\bar{Q}^{1\dot{\alpha}}, \bar{S}^{\dot{\beta}}_1 \right\}
Z_{2} Z_{3} \,=\, \epsilon^{\dot{\alpha}\dot{\beta}}\,
\frac{g^2 N}{4\pi^2} \left[Z_{2}, Z_{3}\right].  
\end{equation}
On the other hand, since the action of $\bar{Q}^1$ on $Z_{2}
Z_{3}$ is zero both at classical level and at order $g^2$,
where it is forbidden by the $SU(4)$ symmetry, the left hand side of
the previous equation reduces to the action of
$\bar{Q}^{1\dot{\alpha}} \bar{S}^{\dot{\beta}}_1$ on the pair
  of fields, which can be computed via~\eqref{s1},
  giving exactly the right hand side of~\eqref{eq:qs}.

We can now use~\eqref{s1} and~\eqref{s2} to compute the first quantum 
correction to the highest weight state~\eqref{ng}. Again for the sake
of concreteness, let us focus on the superconformal charge $\bar
S_1$. By using~\eqref{s1} on the operator~\eqref{ng}, one can see that
the variations involving $\bar Z_1$ yields
\begin{equation}\label{var1ng1}
-\frac{\ii g N}{8 \pi^2} \sqrt{\frac{8 \pi^2}{N}}
\sqrt{\frac{N_0^{-J-1}}{(J + 3)}} 
\sum_{p=0}^{J-1}\, \left(\cos\frac{\pi
  n(2p+3)}{J+3}-\cos\frac{\pi n(2p+5)}{J+3}\right)
\tr{Z_1 Z^p \bar{\psi}^{\dot\alpha}_3 Z^{J-1-p}}~,
\end{equation}
where we have explicitly written one of the $N_0$ factors, because the
resulting operators has only $J+1$ fields, one less in comparison to
the original operator~\eqref{ng}. Eq.~\eqref{var1ng1} can be rewritten
as follows
\begin{equation}\label{var1ng2}
-\ii \,\frac{ g\sqrt{N}}{\sqrt{2} \pi} \sqrt{\frac{N_0^{-J-1}}{(J + 3)}} 
 \sin\frac{\pi n}{J+3} \sum_{p=0}^{J-1}\, \sin\frac{\pi n(2p+4)}{J+3} 
\tr{Z_1 Z^p \bar{\psi}^{\dot\alpha}_3 Z^{J-1-p}}~.
\end{equation}
From~\eqref{s1} it is clear that the action of any (1-loop corrected)
superconformal charge on two identical scalar is trivial. The action
of $\bar S_1$ on the couple $Z Z_2$ yields similar terms involving
$\bar\psi_4$, instead of $\bar\psi_3$.

Exactly as it happened in the string theory computation, we can cancel
these order $g$ contributions against the classical variation of a
term containing two fermionic impurities, but which is suppressed by
an explicit factor of $g$. The result in~\eqref{var1ng2} suggests to
consider the following form for the highest weight state
\begin{eqnarray}
  \label{BMNhws2}
  \mathcal{O}^J_n \,& = &\, \sqrt{\frac{N_0^{-J-2}}{(J + 3)}}
  \sum_{i=1}^3 \sum_{p=0}^J 
  \cos{\frac{\pi n (2 p + 3)}{J + 3}}\tr{ Z_i Z^p \bar{Z}_i Z^{J-p}}
  \\ \nonumber  & + &
  \frac{g \sqrt{N}}{4 \pi}
  \sin\frac{\pi n}{J+3}  
  \; 
  \sqrt{\frac{N_0^{-J-1}}{(J + 3)}}
  \sum_{p=0}^{J-1}
  \sin\frac{\pi n (2p+4)}{J+3}
  \tr{\psi^{1\alpha} Z^p \psi^2_\alpha Z^{J-1-p}}
  \\ \nonumber  & - & 
  \frac{g \sqrt{N}}{4 \pi}
  \sin\frac{\pi n}{J+3}  
  \; 
  \sqrt{\frac{N_0^{-J-1}}{(J + 3)}}\sum_{p=0}^{J-1}
  \sin\frac{\pi n (2p+4)}{J+3}
  \tr{\bar\psi_{3\dot\alpha} Z^p \bar\psi_4^{\dot\alpha} Z^{J-1-p}}~.
\end{eqnarray}      
The coefficients in the second and third line have been fixed in order
to satisfy~\eqref{si}. In fact, we have seen that there is a quantum
contribution from the first line summarized in~\eqref{s1}, and there
are classical contributions from the new terms summarized
in~\eqref{s2}. If we keep focusing on $\bar S_1$,~\eqref{var1ng2}
summarizes the quantum contribution from the planar action on the $Z$
and $Z_1$ which is canceled by the classical variation of $\bar\psi_4$
in the last line. Similar computations show that this pattern applies
also to the action of the other conformal supercharges and to the
action on the other (couple of) fields. Notice also that the
contribution of the boundary terms of \eqref{ng} (\emph{i.e.}~those
with $p=0$ and $p=J$) due to the action of $\bar{S}_1$ on
$Z_1\bar{Z_1}$ and $\bar{Z_1}Z_1$ sum to zero. In fact, the diagrams
involving respectively the pairs $Z_i\bar{Z_i}$ and $\bar{Z_i}Z_i$
come with the same phase factor but opposite sign due to \eqref{s1}.

We close this section with some comments regarding the form of the
primary operator we have derived. Clearly the result~\eqref{BMNhws2}
requires to have $J\geq 1$; if $J=0$, then the highest weight state is
the standard Konishi operator and no mixing is present since it is not
possible to build other $SU(4)$ scalars with the same free scaling
dimension. A second observation is that in the two-point function
$\langle\bar{\mathcal{O}}^J_n \mathcal{O}^J_n\rangle$ there is no
overlap at one loop between the leading and the subleading term of
\eqref{BMNhws2}. In fact the first possible diagram involves three
Yukawa couplings and therefore it is of order $g^3$. Thus,
eq.~\eqref{BMNhws2} can be used as a non trivial test for $H_3$, the
cubic term of the Hamiltonian for the full $PSU(2,2|4)$ theory, which
has not been computed yet. $H_3$ should capture the mixing in
\eqref{BMNhws2} for $p=0$. In order to capture the terms with $p\geq
1$ one would need to compute the two-point function (or the
Hamiltonian) at higher orders in $g$. Moreover, in the large $J$ limit
the corrections to the naive form of the primary, although present, do
not alter the anomalous dimension calculated
in~\cite{Santambrogio:2002sb}. This can be easily seen since any
contribution to the 2-point correlator of the primary coming from
terms like $\langle\tr{\phi^{\ii}Z^l\phi^{\ii}Z^{J-l}} \tr{\psi
  Z^l\psi Z^{J-1-l}}\rangle$ should involve both impurities and is
thus suppressed in the large $J$ limit. Also in the computation of the
1-loop anomalous dimension, all the corrections terms in~\eqref{s1}
can be neglected even at finite $J$. Finally from~\eqref{BMNhws2} it
is possible to derive the descendant operators , either by the same
method employed here for the primary, with the only difference that
one has to use the supersymmetry current instead of the superconformal
one, or by using the Konishi anomaly~\cite{Bianchi:2003eg}.

\sect{Comparison with the predictions from string theory}
\label{sec:comp}

In section~\ref{FT} we saw that the mixing between scalar and fermion
impurities in the $\mathcal{N}=4$ SYM primary we considered mirrors exactly the
patterns we found in the correspondent string theory
computation. Actually, if we expand the string theory highest weight
state in~\eqref{shws} in powers of $(\mu\alpha)^{-1}$, the first
subleading term, quadratic in the fermionic oscillators, matches
exactly with the large $J$ expansion of the second and third line
in~\eqref{BMNhws2}. In particular, up to order $\frac{1}{\mu\alpha} =
\sqrt{\lambda'}$, the string state~\eqref{shws} becomes:
\begin{equation}
  \label{hwsmu}
  |n\rangle \approx \frac{1}{4}
  \left[
    {a^\dagger}_n^{i'}
    {a^\dagger}_n^{i'}
    \,+\, 
    {a^\dagger}_{-n}^{i'}
    {a^\dagger}_{-n}^{i'}
    + n\sqrt{\lambda'}
    \left( 
      {b^\dagger}_{-n} 
      \mathbb{P}^+
      {b^\dagger}_n
      -
      {b^\dagger}_{-n} 
      \mathbb{P}^-
      {b^\dagger}_{n}
    \right)
  \right]
  |\alpha\rangle~.
\end{equation}
We can translate the bosonic and fermionic contributions of this
formula into the corresponding BMN operators. In this way, we can
read from~\eqref{hwsmu} a prediction for the BMN limit of the first
and second terms of the quantum corrected SYM highest weight operator.
Let us start from the term in~\eqref{hwsmu} with the bosonic
oscillator. According to the standard PP-wave/BMN dictionary we have
  \begin{equation}
  \label{eq:a1}
  \frac{1}{4}\left[
    {a^\dagger}_n^{i'}
    {a^\dagger}_n^{i'}
    \,+\, 
    {a^\dagger}_{-n}^{i'}
    {a^\dagger}_{-n}^{i'}
  \right]|\alpha\rangle
  \,\leftrightarrow\, \mathcal{O}^{(0) J}_n ~,
\end{equation}
where $\mathcal{O}^{(0) J}_n$ is defined in~\eqref{ng}. The string
state on the l.h.s. is normalized to one as it is the tree-level
2-point function of the corresponding gauge theory operator
\begin{equation}\label{nop}
\langle\bar{\mathcal{O}}^{(0) J}_n(x) \mathcal{O}^{(0) J}_n(y) \rangle
= \frac{(-1)^{J+2}}{(x-y)^{2(J+2)}}, 
\end{equation}
The term with the fermionic oscillators in~\eqref{hwsmu} correspond to
a gauge theory operator with two spinors:
\begin{equation}
  \label{eq:a2}
  \frac{1}{2}
  \left[
    {b^\dagger}_{-n}
    \mathbb{P}^+
    {b^\dagger}_n  
  \right]|\alpha\rangle
  \,\leftrightarrow\,
  \mathcal{O}^{(1) J}_n = \frac{1}{2}
  \sqrt{\frac{N_0^{-J-1}}{J+1}}
  \sum_{p=0}^{J-1}\sin{\frac{\pi n(2p+2)}{J+1}}
  \tr{\psi^{1 \alpha} Z^p \psi^2_{\alpha} Z^{J-p-1}}
\end{equation}
and the similar formula relates the $\mathbb{P}^-$ projection of the
string state to a gauge theory operator with the spinors $\bar\psi_3$
and $\bar\psi_4$. Also in this case the normalizations have been fixed
by requiring that the SYM tree-level 2-point function takes the
canonical form~\eqref{nop} and that the norm of the string state is
one.

Then rewriting $\sqrt{\lambda'}=\frac{\sqrt{N} g_{YM}}{J}$ we get that
the gauge theory operator corresponding to the string
state~\eqref{hwsmu} is:
\begin{align}
  {(\mathcal{O}_{st})}^J_n & \,=\,
   \sqrt{\frac{N_0^{-J-2}}{J+3}} 
    \sum_{p=0}^J\cos{\frac{\pi n(2p+3)}{J+3}}
    \tr{Z_i Z^p \bar{Z}_i Z^{J-p}} + \\ \nonumber
  &  \,+\, \frac{g \sqrt{N} n}{4J} \sqrt{\frac{N_0^{-J-1}}{J+1}}  
    \sum_{p=0}^{J-1}\sin{\frac{\pi n(2p+2)}{J+1}}  
    \tr{\psi^{1 \alpha} Z^p \psi^2_{\alpha} Z^{J-p-1}} + \\ \nonumber
  &  \,-\, \frac{g \sqrt{N} n}{4J} \sqrt{\frac{N_0^{-J-1}}{J+1}}
    \sum_{p=0}^{J-1}\sin{\frac{\pi n(2p+2)}{J+1}}  
    \tr{\bar\psi_{3 \dot\alpha} Z^p \bar\psi_4^{\dot\alpha} Z^{J-p-1}} ~.
\end{align}
In the large $J$ limit this result agrees with~\eqref{BMNhws2} derived in
section~\eqref{FT} by using perturbative field theory.

\sect{Conclusions}

In this paper, we derived the expression of the two impurities highest
weight state for type IIB string theory on the maximal supersymmetric
PP-wave background. Our string result is valid at tree-level, but is
exact in $\mu\alpha$. Moreover, we used the superconformal properties
of $\mathcal{N}=4$ SYM to derive the two-impurity highest weight state
of the $SU(2,2|4)$ superconformal algebra at finite $J$ and up to the
order $g$. In both cases, the naive form of the state is corrected by
a term containing two fermionic impurities. Then we showed that the
large-$J$ limit of the gauge theory highest weight state matches with
the strong curvature expansion of the string state to the same order
in perturbation theory. This precise agreement is partially
unexpected, as the BMN string theory is valid in the limit
$\lambda,J\to\infty$ with $\lambda'$ fixed, while perturbative gauge
theory computations require $\lambda\ll 1$. We do not know any clear
reason why these two different scaling limits should match exactly.
However this happens for the planar anomalous dimensions up to order
$g^6$~\cite{Gross:2002su,Eden:2004ua}. From our results it follows
that also the form of the operators, not just their dilatation
eigenvalue, matches up to order $g$.

It would be certainly interesting to check whether the agreement
between the string and the field theory highest weight states survives
at the next order ${\cal O}(g^2)$. In order to check this, one would
need to know more about the quantum corrected form of the conformal
supercharges. This can done by following the approach discussed in
section~\ref{FT} and by deriving the quantum corrected action of the
$S$ and $\bar{S}$ on fermion and vector fields. For instance, there
will be certainly a contribution of the form $\bar{S} \psi \Phi \sim
\slash \!\!\!\partial \Phi$. This will induce on the field theory side
the same pattern we have seen in section~\ref{ST} and thus we expect
that, at the order $g^2$, also derivative impurities appear in the
explicit form of the field theory highest weight state.  On the other
hand our field theory result is exact in $J$. It would be interesting
to consider the subleading corrections in the large-$J$ limit. These
should be captured by the near-BMN limit of the $AdS_5\times S^5$
string theory which has been thoroughly studied~\cite{Callan:2003xr}.
In particular, the form of the string supercharges in the near-BMN
limit is known in the literature~\cite{Arutyunov:2006ak}. Taking into
account these corrections on the string side would allow another
comparison with our gauge theory result.  Another possible extension
of the approach proposed here is to consider the non-planar action of
the conformal supercharges. It would be interesting to see whether
this is sufficient to reproduce the $1/N$ mixing obtained
in~\cite{Beisert:2002bb,Constable:2002vq} by using the standard
approach of the orthonormalization of the 2-point correlators.

As pointed out in section~\ref{FT}, the corrections we derived to the
field theory operators are not needed in the computation of the ${\cal
O}(g^2)$ anomalous dimensions and also can be neglected in the BMN
limit at all orders. However, we expect that they play a crucial role
in the computation of three and higher point correlators. It would be
certainly interesting to see this in some explicit examples. This
would provide a systematic basis to study the correspondence between
BMN string theory and gauge theory in presence of fermionic impurities
and generalize the results of~\cite{Georgiou:2004ty,Dobashi:2006fu}.

\vspace{1cm}

\noindent {\large {\bf Acknowledgments}}

\vspace{3mm}

\noindent
We wish to thank Andreas Brandhuber, Tom Brown, Paul Heslop, Sanjaye
Ramgoolam and Gabriele Travaglini for useful discussions and comments.
V.~G. and R.~R. wish to thank Paolo Di Vecchia and Alessandro Tanzini
for collaboration on related topics. The work of G.~G. is supported by
STFC through a Postdoctoral Fellowship and that of V.~G. is supported
by the Foundation Boncompagni-Ludovisi.  This work is
partially supported by the EC Marie Curie Research Training Network
MRTN-CT-2004-512194.

\appendix

\sect{$\mathcal{N} =4$ SYM conventions}
\label{sec:GTconv}

The Lagrangian and the supersymmetry variations of the four
dimensional $\mathcal{N} = 4$ SYM can be derived by dimensional
reduction from the ten dimensional $N=1$ SYM
theory~\cite{Brink:1976bc}.  Here we recall the main steps of this
derivation mainly with the aim of setting up some notations that are
useful for building the dictionary between BMN string states and gauge
theory operators.

The ten-dimensional action is 
\begin{equation}
  S_{10}=\int d^{10}x 
  \tr{-\frac{1}{2} F_{MN}F^{MN} + 
    \ii \bar{\lambda} \Gamma^M D_M \lambda}
\end{equation}
We adopt the ``mostly-minus'' metric $(+,-,\ldots,-)$ and the
following conventions for the gauge group generators:
\begin{equation}\label{colcon}
  \tr{T^a T^b} \,=\, \frac{\delta^{ab}}{2}~,~~~
  \left[T^a , T^b \right] \,=\, i f^{abc} T^c~,~~~
  (T^a)^i_j(T^a)^k_l \,=\, 
  \frac{1}{2}\left(
    \delta^i_l \delta^k_j 
    - \frac{1}{N} \delta^i_j \delta^k_l 
  \right)
\end{equation}

A useful representation of the the ten dimensional Gamma matrices with
mostly minus signature is
\begin{equation}
\Gamma^\mu = 1_8 \otimes \gamma^\mu ~~,~~~
\Gamma^{i+3} = \sigma^1 \otimes \eta^i \otimes \gamma^5~~,~~~
\Gamma^{i+6} = - \sigma^2 \otimes \bar\eta^i \otimes \gamma^5~,
\end{equation}
where $1_n$ is the $n \times n$ identity matrix, the $\gamma^\mu$'s
are the standard four dimensional gamma matrices in the Weyl
representation
\begin{equation}
  \label{gammaM}
  \gamma^\mu = \left(
    \begin{array}{cc}
      0 & \sigma^\mu \\ \bar\sigma^\mu & 0 
    \end{array}\right)~~~\mbox{and}~~~~
\gamma^5= \ii \prod_{j=0}^3 \gamma^j = \left(
    \begin{array}{cc}
      -1 & 0 \\ 0 & 1 
    \end{array}\right),
\end{equation}
with $\sigma^0=\bar\sigma^0=1_2$ is, while $\sigma^i=-\bar\sigma^i$
are the standard Pauli matrices.

The $\sigma$-matrices satisfy the following relation:
\begin{equation}
  \label{eq:twosigma}
  \sigma^\mu_{\alpha\dot{\alpha}}\,
  \sigma^\nu_{\beta\dot{\beta}}\, 
  \epsilon^{\dot{\alpha}\dot{\beta}} 
  \,=\,
  -\eta^{\mu\nu}\epsilon_{\alpha\beta} \,+\, 
  2 \sigma^{\mu\nu}_{\alpha\beta}
\end{equation}
where we have defined $\sigma^{\mu\nu}_{\alpha\beta} = \frac{1}{4}\left(
\sigma^{\mu}_{\alpha\dot{\alpha}}\, \sigma^{\nu}_{\beta\dot{\beta}}\,
-\,\sigma^{\nu}_{\alpha\dot{\alpha}}\, \sigma^{\mu}_{\beta\dot{\beta}}
\right)
\epsilon^{\dot{\alpha}\dot{\beta}}$, and $\epsilon_{12}=\epsilon^{21}=1$. 

Finally $\eta^i\,,~\bar\eta^i$ are the 't~Hooft matrices
\begin{subequations}
  \label{eq:tm}
  \begin{gather}
    \eta^i_{AB} \,=\, \delta_{iA} \delta_{B4}
    - \delta_{iB} \delta_{A4} 
    + \epsilon_{iAB4},
    \label{eta}\\
    \bar{\eta}^i_{AB} \,=\, \delta_{iA} \delta_{B4}
    - \delta_{iB} \delta_{A4} 
    - \epsilon_{iAB4},
    \label{bareta}
  \end{gather}
\end{subequations}
which satisfy 
$\eta^i \eta^j = -\delta^{ij} 1_4 - \epsilon^{ijk}\eta^k$
and
$\bar{\eta}^i \bar{\eta}^j = -\delta^{ij} 1_4 + \epsilon^{ijk}\bar{\eta}^k$.

In this representation we have $\Gamma^{11}=\sigma^3\otimes 1_4\otimes
\gamma^5$. The gaugino of the ten dimensional theory $\lambda$ is a
Majorana-Weyl spinor ($\Gamma^{11} \lambda=\lambda$) which, with the
conventions above can be express in term of the four Weyl plus four
anti-Weyl spinors of the four dimensional theory, $\psi_{\alpha}^A$
and $\bar{\psi}^{\dot\alpha}_A$ respectively:
\begin{equation}
  \label{10dgaugino}
\lambda^t= [(0,\bar\psi^{\dot\alpha}_{A=1}),\ldots,
(0,\bar\psi^{\dot\alpha}_{A=4}),(\psi_{\alpha}^{A=1},0),\ldots,
(\psi_{\alpha}^{A=1},0)]~,
\end{equation}
The action of the chirality matrix gives $\gamma^5\,
{}^t(0,\bar\psi^{\dot{\alpha}})={}^t(0,\bar\psi^{\dot{\alpha}})$ and
$\gamma^5\,{}^t(\psi_\alpha,0)=-{}^t(\psi_\alpha,0)$.  

The index $A$ rotates into representations of the internal $SU(4)$
R-symmetry of the four dimensional $\mathcal{N} = 4$ SYM theory. In
particular, the Weyl spinor $\psi_\alpha^A$ transform in the
fundamental representation, while their conjugates
$\bar{\psi}_A^{\dot{\alpha}}$ transform in the antifundamental one.

The six scalar fields arising from the internal components of the
gauge field can be organized into the components $\Phi_{AB}$ of a
tensor in the antisymmetric representation of $SU(4)$:
\begin{equation}
  \label{PhiAB}
  \Phi_{AB} = \frac{1}{2\sqrt{2}} \sum_{j=1}^3\left[
    A_{j+3} \eta^j_{AB} + \ii A_{j+6} \bar{\eta}^j_{AB}\right]~.
\end{equation}

With these conventions, the dimensionally reduced Lagrangian is
\begin{multline}
  \label{n4l}
  L  =  \tr{- \frac{1}{2} F_{\mu \nu} F^{\mu \nu} +
    2 D_{\mu} {\Phi}_{AB}  D^{\mu} \Phi^{AB}   + 2 i \psi^{\alpha A}
    \sigma_{\alpha \dot{\alpha}}^{\mu} 
    (D_{\mu} {\bar{\psi}}^{\dot{\alpha}}_{A})
    +\\ 
    2 g^2 [\Phi^{AB}, \Phi^{CD}] [{{\Phi}}_{AB},
    {{\Phi}}_{CD}] -  g 2 \sqrt{2} \left([\psi^{\alpha A},
      {{\Phi}}_{AB}]
      \psi_{\alpha}^{B} - [{\bar{\psi}}_{\dot{\alpha} A}, \Phi^{AB}]
      {\bar{\psi}}^{\dot{\alpha}}_{B}\right)}
\end{multline}
where the scalar fields with upper indices
$\Phi^{AB}$ are defined as follow:
\begin{equation}
  \label{eq:barPhi}
  {\Phi}^{AB} = \frac{1}{2} \epsilon^{ABCD} \Phi_{CD} \equiv
  {\Phi}^*_{AB},
\end{equation}
and the covariant derivative is $D_\mu \phi = \partial_\mu \phi - i g
[A_\mu,\phi]$.

Out of \eqref{n4l} we read the  Minkowskian free scalar propagator:
\begin{equation}
  \label{eq:mfsp}
  \langle
  Z_i^a(x)\,\bar{Z}_j^b(y)
  \rangle
  \,=\,  
  \delta_{ij}
  \delta^{ab}
  \Delta_{xy}
\qquad 
\Box_x\Delta_{xy} = - \ii \delta^4(x-y)
\end{equation}
and the free fermionic one:
\begin{equation}
  \label{eq:mffp}
  \langle  
  \psi^{A a}_\alpha(x) 
  \bar{\psi}^b_{\dot{\alpha} B}(y)
  \rangle \,=\,
  \ii 
  \delta^{ab}
  \delta^A_B
  \sigma^\mu_{\alpha\dot{\alpha}}
  \partial^x_\mu\Delta_{xy}
\quad\Rightarrow\quad
  \langle  
  \bar{\psi}^{\dot{\alpha} a}_{A}(x) 
  \psi^{\alpha B b}(y)
  \rangle \,=\,
  \ii 
  \delta^{ab}
  \delta^A_B
  \bar{\sigma}^{\mu\,\dot{\alpha}\alpha}
  \partial^x_\mu\Delta_{xy},
\end{equation}
where $\Delta_{xy} = - \frac{1}{4\pi^2\,(x-y)^2}$, and the
$\sigma$-matrices are defined after formula \eqref{gammaM}.

The 10D supersymmetry transformation, $\delta A_M = \ii
\bar\xi \Gamma_M \lambda$ and $\delta \lambda = \frac{1}{4} [\Gamma^M,
\Gamma^N]F_{MN} \xi$, decompose as follows:
\begin{subequations}\label{susyt}
  \begin{align}
\nonumber \bar\xi \,=\; &
[(\xi^{\alpha\,A=1},0),\ldots,(\xi^{\alpha\,A=4},0),
(0,\bar\xi_{\dot\alpha\,A=1}),\ldots,(0,\bar\xi_{\dot\alpha\,A=4})] &
\\ \label{a131}
\delta \Phi_{AB} \,=\, & \frac{\ii}{\sqrt{2}} \left[
\epsilon_{ABCD}\,\xi^{C\,\alpha}\,\psi^D_\alpha \,-\,
\bar{\xi}_{A\,\dot{\alpha}} \bar{\psi}_{B}^{\dot{\alpha}} \, + \,
\bar{\xi}_{B\,\dot{\alpha}} \bar{\psi}_{A}^{\dot{\alpha}} \right] 
\\\label{a132}
\delta \psi^A_\alpha \,=\, & \sigma^{\mu \nu \phantom{\alpha}
\beta}_{\phantom{\mu\nu}\alpha}\, \xi^A_\beta\, F^{\mu \nu} \,+\,
2\sqrt{2}\sigma^\mu_{\phantom{\mu}\alpha \dot{\alpha}}\,
\bar{\xi}^{\dot{\alpha}}_B\, D_\mu {\Phi}^{AB}\, -\,4 \ii g \left[
{\Phi}^{A C}, \Phi_{CB}\right]\, \xi_\alpha^{B}
\\ \label{a133}
\delta
\bar{\psi}_A^{\dot{\alpha}} \,=\, & \bar{\sigma}^{\mu \nu
\,\dot{\alpha}}_{\phantom{\mu\nu\beta}\dot{\beta}}\,
\bar{\xi}_A^{\dot{\beta}}\, F^{\mu \nu} \,-\,
2\sqrt{2}\bar{\sigma}^{\mu\,\dot{\alpha} \alpha}\, \xi^B_\alpha\, D_\mu
\Phi_{AB}\,-\, 4 \ii g \left[ \Phi_{A C}, {\Phi}^{CB}\right]\,
\bar{\xi}^{\dot{\alpha}}_{B}
\\ \label{a134}
\delta A_\mu \,=\,& \ii\left(
\sigma_{\mu\,\alpha\dot{\alpha}} \xi^{\alpha A}
\bar{\psi}^{\dot{\alpha}}_A \,+\,
\bar{\sigma}_\mu^{\,\dot{\alpha}\alpha} \bar{\xi}_{\dot{\alpha} A}
\psi_\alpha^A \right)
  \end{align}
\end{subequations}

The supercurrent associated to the invariance of the ten-dimensional
theory under supersymmetry transformations is:
\begin{equation}
  \label{eq:10dcurrent}
  Q^M = \frac{\ii}{2}[\Gamma^R,\Gamma^N] \tr{F_{RN} \Gamma^M \lambda} , 
  \quad M = 0, \ldots, 9
\end{equation}

After dimensional reduction to $D=4$, we get the following
supersymmetric current for $\mathcal{N}=4$ SYM:
\begin{equation}
  \label{eq:4dcurrmu}
  Q^\mu = {}^{t}\left[(Q^\mu_{\alpha A=1}, 0),\ldots(Q^\mu_{\alpha
      A=4},\, 0), \,(0,\, \bar{Q}^{\mu\,\dot{\alpha} A=1}),\ldots(0,\,
    \bar{Q}^{\mu\,\dot{\alpha} A=4})\right]
\end{equation}
with 
\begin{subequations}
  \begin{align}
    Q^\mu_{\phantom{\mu}\alpha A} & =
    2 \ii \tr{(\sigma^{\rho \nu})_\alpha^\beta F_{\rho \nu} 
      \sigma^{\mu}_{\beta\dot{\beta}}\bar{\psi}^{\dot{\beta}}_{A}
      +2\sqrt{2} D_\rho \Phi_{AB} \sigma^\rho_{\alpha\dot{\alpha}} 
      \bar{\sigma}^{\mu\,\dot{\alpha}\beta} \psi_{\beta}^B
      - 4 \ii g [\Phi_{AC},\Phi^{CB}] 
      \sigma^\mu_{\alpha\dot{\alpha}}\bar{\psi}^{\dot{\alpha}}_{B}}\\
    \bar{Q}^{\mu\,\dot{\alpha} A} & =
    2 \ii \tr{(\bar{\sigma}^{\rho \nu})^{\dot{\alpha}}_{\dot{\beta}} F_{\rho \nu} 
      \bar{\sigma}^{\mu\,\dot{\beta}\beta}\psi^{A}_{\beta}
      -2\sqrt{2}D_\rho \Phi^{AB} \bar{\sigma}^{\rho\,\dot{\alpha}\alpha}
      \sigma^\mu_{\alpha\dot{\beta}} \bar{\psi}^{\dot{\beta}}_B
      - 4 \ii g[\Phi^{AC},\Phi_{CB}] 
      \bar{\sigma}^{\mu\dot{\alpha}\alpha}\psi_{\alpha}^B}
  \end{align}
\end{subequations}

On the other hand, the current associated to the superconformal
transformations are obtained first by replacing, in the supersymmetry
variation of a field, the supersymmetry parameters $\xi^{\alpha A}$
and $\bar{\xi}_{\dot{\alpha} A}$ with
$\ii x_{\mu}\bar{\sigma}^{\mu\,\dot{\alpha}\alpha}
\bar{\zeta}^A_{\dot{\alpha}}$ and
$\ii x_{\mu}\sigma^\mu_{\alpha\dot{\alpha}}\zeta^\alpha_A$ respectively,
and then adding all possible $x$-independent terms with the same mass
dimension and the same quantum numbers which are compatible with the
superconformal algebra~\cite{Mehta:1988za}. Out of this process, we
get the following superconformal currents:
\begin{subequations} 
\label{supc} 
\begin{multline}
\label{eq:bscc}
\bar{S}^{\mu \dot\alpha}_{\phantom{\mu} A}  =  
  2  x_\tau (\bar\sigma^\tau)^{\dot\alpha \alpha} 
\tr{(\sigma^{\rho \nu})_\alpha^\beta F_{\rho \nu} 
      \sigma^{\mu}_{\beta\dot{\beta}}\bar{\psi}^{\dot{\beta}}_{A}
      + 2\sqrt{2} D_\rho \Phi_{AB} \sigma^\rho_{\alpha\dot{\beta}} 
      \bar{\sigma}^{\mu\,\dot{\beta}\beta} \psi_{\beta}^B+\\    
      - 4 \ii g [\Phi_{AC},\Phi^{CB}]\sigma^\mu_{\alpha\dot{\beta}}
      \bar{\psi}^{\dot{\beta}}_{B}}  + 8\sqrt{2}  
      \tr{\phi_{AB} (\bar\sigma^{\mu})^{\dot\alpha \alpha} \psi^B_\alpha},
\end{multline}
\begin{multline}
\label{eq:scc}
  {S^\mu}^A_\alpha = 
    2  x_\tau \sigma^\tau_{\alpha \dot{\alpha}}
\tr{(\bar{\sigma}^{\rho \nu})^{\dot{\alpha}}_{\dot{\beta}} F_{\rho \nu} 
      \bar{\sigma}^{\mu\,\dot{\beta}\beta}\psi^{A}_{\beta}
      -2\sqrt{2}D_\rho \Phi^{AB} \bar{\sigma}^{\rho\,\dot{\alpha}\beta}
      \sigma^\mu_{\beta\dot{\beta}} \bar{\psi}^{\dot{\beta}}_B +\\
      - 4 \ii g[\Phi^{AC},\Phi_{CB}] 
      \bar{\sigma}^{\mu\dot{\alpha}\beta}\psi_{\beta}^B} - 8\sqrt{2} 
      \tr{\phi^{AB} \sigma^{\mu}_{\alpha \dot{\alpha}} \bar{\psi}_B^{\dot{\alpha}}}
\end{multline}
\end{subequations}
where the coefficients of the last terms have been fixed by requiring
that $\partial_\mu S^\mu = 0$ and $\partial_\mu \bar{S}^\mu = 0$
on-shell.

We refer to section 3 of \cite{Dolan:2002zh} for the four-dimensional
superconformal algebra closed by the charges associated to the
currents in \eqref{eq:4dcurrmu} and \eqref{supc}, together with the
generator of the conformal algebra in four dimensions.

At this point it is straightforward also to make contact with the
${\cal N}=1$ formalism where the scalars are arranged into three
complex fields 
\begin{eqnarray} \label{comp}
&  Z_i=(A_{i+3}+\ii A_{i+6})/\sqrt{2} &
\\ \nonumber &
  \Phi_{14} = \frac{1}{2} Z_1,\quad
  \Phi_{24} = \frac{1}{2} Z_2,\quad
  \Phi_{34} = \frac{1}{2} Z_3,\quad
  \Phi_{13} = - \frac{1}{2} \overline{Z}_2\,,\quad
  \Phi_{23} = \frac{1}{2} \overline{Z}_1,\quad
  \Phi_{12} = \frac{1}{2} \overline{Z}_3. &
\end{eqnarray}
We select the $U(1) \in SU(4)$ which rotates $Z_3 \doteq Z$ as the BMN
$U(1)$. In order to see the fate of the various spinors in the BMN
limit, it is convenient to compute their charges under the Cartan
generators $(J_{Z_1}^{(s)},J_{Z_2}^{(s)},J_{Z_3}^{(s)})$, with
$J_{Z_i}^{(s)} = {-\ii\Gamma^{i+3} \Gamma^{i+6}}$, that rotate the
complex scalars ($Z_1,Z_2,Z_3)$
\begin{equation}
  \label{cartanqn}
\psi^1 \to (-\frac{1}{2},\frac{1}{2},\frac{1}{2})~~,~~~
\psi^2 \to (\frac{1}{2},-\frac{1}{2},\frac{1}{2})~~,~~~
\psi^3 \to (\frac{1}{2},\frac{1}{2},-\frac{1}{2})~~,~~~
\psi^4 \to (-\frac{1}{2},-\frac{1}{2},-\frac{1}{2}) ~.
\end{equation}
Of course $\bar\psi_A$ has the opposite assignments. The
supersymmetries $Q_{\alpha\,A=1,2}$ have the same
chirality both in in the space-time and in the $Z_1,Z_2$ directions
and they correspond in the PP-wave string theory to the kinematical
generators $\mathbb{P}^+Q^+/2 $. Similarly the supersymmetries
$\bar Q^{\dot\alpha\,A=3,4}$ correspond in to the other kinematical
generators $\mathbb{P}^-\bar{Q}^+/2 $. This assignment is also consistent
with the fact that these supersymmetries act non-trivially on $Z_3$
and thus generate the BPS multiplet starting from the
operator $\tr{Z^J}$ in field theory or the vacuum state
$|p^+\rangle$ in the string theory language. The other supersymmetry
variations have opposite chirality in the space-time and in the
internal $Z_1,Z_2$ plane and thus must correspond to the dynamical
supercharges $Q^-$ and $\bar{Q}^-$.

\sect{A useful integral}
The computation of the one-loop correlators in section \ref{FT}
involve the following integration (let us remember that $\sigma^\mu =
(1_2,\sigma^i_{(P)}), i=1,2,3$):
\begin{align}
I^M(x_1; x_2, x_3) \,\doteq\, & 
\sigma^\mu_{\alpha\dot{\alpha}} \sigma^\nu_{\beta\dot{\beta}}
\epsilon^{\dot{\alpha} \dot{\beta}}
\int d^4x \frac{1}{(x_1 - x)^2} 
\partial_\mu^{x_2} \frac{1}{(x_2 - x)^2} 
\partial_\nu^{x_3} \frac{1}{(x_3 - x)^2} \notag\\
\,=\, & -\ii 4\pi^2
\sigma^\mu_{\alpha\dot{\alpha}} \sigma^\nu_{\beta\dot{\beta}} 
\epsilon^{\dot{\alpha} \dot{\beta}}
\frac{{x_{12}}_\mu {x_{13}}_\nu}{x_{12}^2 x_{13}^2 x_{23}^2}
  \label{eq:IM}
\end{align}

The last identity follows by first analytically continuing the
integral to Euclidean spacetime, and then computing separately the
symmetric and antisymmetric components in the spacetime indices. The
computation of the symmetric part is straightforward, while the
antisymmetric one is evaluated by connecting it to another conformal
integral, this time in $d=6$ (Euclidean)
dimensions~\cite{Usyukina:1994iw}:
\begin{equation}
  \label{eq:I6}
  I_6(y_1,y_2,y_3) \doteq 
  \int d^6y \left[
\frac{1}{(y_1-y)^2\,(y_2-y)^2\,(y_3-y)^2}
\right]
= \pi^3 \frac{1}{y_{12}^2\,y_{13}^2\,y_{23}^2}.
\end{equation}

Introducing $x^0_M \doteq - \ii x^4_E$, we can rewrite  \eqref{eq:IM} as
\begin{align}
  \label{eq:IE}
  I^M(x_1; x_2, x_3) = & 
- \ii \sigma^m_{\!(E)\,\alpha\dot{\alpha}} \sigma^n_{\!(E)\,\beta\dot{\beta}}
\epsilon^{\dot{\alpha} \dot{\beta}} 
\int_E d^4x \frac{1}{(x_1 - x)^2} 
\partial_m^{x_2} \frac{1}{(x_2 - x)^2} 
\partial_n^{x_3} \frac{1}{(x_3 - x)^2} \notag \\
\doteq &
- \ii \sigma^m_{\!(E)\,\alpha\dot{\alpha}} \sigma^n_{\!(E)\,\beta\dot{\beta}}
\epsilon^{\dot{\alpha} \dot{\beta}} 
I^E_{m n}(x_1; x_2, x_3)
\end{align}
where we have defined $\sigma^m_{\!(E)} = (\ii \sigma^j_{(P)}, -1_2),
j=1,2,3$.

Remembering that $\sigma^m_{\!(E)\,\alpha\dot{\alpha}}
\sigma^n_{\!(E)\,\beta\dot{\beta}} \epsilon^{\dot{\alpha} \dot{\beta}}
= - \delta^{m n}\epsilon_{\alpha\beta} + 2 \sigma^{m n}_{\alpha
  \beta}$, we can decompose the integral into its symmetric plus
antisymmetric part. The symmetric part has been computed
in~\cite{Dobashi:2006fu}, the result being $\delta^{m n}I^E_{m n}(x_1;
x_2, x_3) = 4 \pi^2 \frac{{x_{12}}_m {x_{13}}_n \delta^{m n}}
{x_{12}^2 x_{13}^2 x_{23}^2}$.  The antisymmetric one can be evaluated
via a comparison with the integral in equation \eqref{eq:I6}, after
writing both of them in term of Schwinger parameters. We get:
\begin{equation}
  I_6 = \int d^6 y\,
  \int_0^\infty 
  \left(
    \prod_{i=1}^3 \frac{d\alpha_i\,\alpha_i}{\Gamma(2)}
  \right)   
  e^{-\sum_i\alpha_i(y_i - y)^2},
\end{equation}
which, after the Gaussian integration over $y$ and after introducing
the new integration variables 
\begin{equation}
\label{eq:vc}
 \hat{\alpha}_i =
\frac{\alpha_i}{T}~,~~ i= 1,2,3, \quad\text{with}~~ T=\sum_i\alpha_i
\end{equation}
can be rewritten as:
\begin{equation}
  \label{eq:I6fin}
  I=\pi^3 \int_0^\infty d T\,T^2
  \int_0^1 \left(
    \prod_i d \hat{\alpha}_i\,\hat{\alpha}_i
  \right)   
  \d{1-\sum_i\hat{\alpha}_i}
  e^{- T \sum_{i<j} \hat{\alpha_i} \hat{\alpha_j}\,y_{ij}^2}.
\end{equation} 

On the other hand, in term of Schwinger parameters $I^E_{[m n]}(x_1;
x_2, x_3)$ becomes:
\begin{equation}
  I^E_{[m n]}(x_1; x_2, x_3) \,=\,
  \int d^4u \partial_{[m}^{x_2} \partial_{n]}^{x_3} 
  \int_0^\infty d\alpha_i e^{- \sum_i \alpha_i (x_i - x)^2}.
\end{equation}
The integrand can be simplified by taking the derivatives and
exploiting the antisymmetry in the spacetime indices. Then, performing
the change of variables in \eqref{eq:vc} and the Gaussian integration
over $x^m$ we get:
\begin{equation}
  \label{eq:Iasy}
  I^E_{[m n]}(x_1; x_2, x_3) \,=\,
  4 \pi^2 {x_{12}}_{[m} {x_{13}}_{n]}\int_0^\infty d T\,T^2 
  \int_0^1 \left(\prod_{i=1}^3 d \hat{\alpha}_i \,\alpha_i\right)
  \d{1-\sum_i\hat{\alpha}_i}
  e^{-T\sum_{i<j}\hat{\alpha}_i \hat{\alpha}_j x_{ij}^2}.
\end{equation}

The direct comparison between \eqref{eq:Iasy} and \eqref{eq:I6fin}
shows immediately:
\begin{equation}
  \label{eq:asym}
  I^E_{[m n]}(x_1; x_2, x_3) \,=\, 
\frac{1}{\pi} I_6(y_1,y_2,y_3)\vert_{y_{ij}^2 \doteq x_{ij}^2} 
\,=\, 4 \pi^2 \frac{{x_{12}}_{[m} {x_{13}}_{n]}}{x_{12}^2 x_{13}^2 x_{23}^2}
\end{equation}

Putting together the result in \eqref{eq:asym} together with the
symmetric part computed before, and rotating back the result to
Minkowskian space-time we finally get:
\begin{equation}\label{integral}
  I^M(x_1; x_2, x_3) \,=\, 
- \ii 4 \pi^2 \sigma^\mu_{\alpha\dot{\alpha}} \sigma^\nu_{\beta\dot{\beta}}
\epsilon^{\dot{\alpha} \dot{\beta}} 
\frac{{x_{12}}_\mu {x_{13}}_\nu}{x_{12}^2 x_{13}^2 x_{23}^2}
\end{equation}

\end{document}